\documentclass[aps,prd,twocolumn,showpacs,superscriptaddress,nofootinbib]{revtex4-1}

\usepackage{amsmath}
 
\usepackage[latin1]{inputenc}
\usepackage{txfonts}
\usepackage{hyperref}
\usepackage[pdftex]{graphicx}

\usepackage{xcolor}

\begin{document}

\title{Evolving scale-free networks  and node-based
  random edge deletion }
\author{Everton R. Constantino} 
\author{Alberto Saa} 
\email{asaa@ime.unicamp.br}
\affiliation{
Department of Applied Mathematics, 
 University of Campinas,  13083-859 Campinas, SP, Brazil.}

\date{\today}

\begin{abstract}
We investigate a growing network model that combines preferential and uniform attachment with two distinct mechanisms of edge deletion. In addition to the usual uniform random edge deletion, we introduce a novel node-based  rule in which uniformly chosen non-isolated nodes lose one of their incident edges. This mechanism differs fundamentally from uniform edge deletion and leads to a nonlinear evolution for the stationary degree distribution due to the nonlinear dependence on the fraction of isolated nodes. We solve the general problem in the stationary regime and obtain closed-form expressions for the degree distribution in terms of hypergeometric and confluent hypergeometric functions. Depending on the balance between attachment and deletion rates, three asymptotic regimes  for the degree distribution arise: power-law, exponential, and a critical regime characterized by a stretched exponential decay.  We show that the node-based edge deletion mechanism is less likely to disrupt the scale-free (power-law) regime than the uniform edge deletion. Moreover, we also demonstrate that the precise balance between preferential attachment and node-based deletion can transform a scale-free network into a critical one with stretched exponential decay.  Large-scale numerical simulations exhibit  excellent agreement with the theoretical predictions.
 \end{abstract}

\maketitle
 
{\bf Scale-free networks are a paradigmatic structure underlying a myriad of systems, from the internet to social and information networks. This paper studies how such networks evolve when growth is combined not only with the addition of new nodes and links, but also with two different kinds of link loss, including a new ``node-based'' edge-deletion mechanism, in which randomly chosen active nodes lose one of their connections, besides the usual random deletion of edges. Depending on the balance between attachment and deletion, the network can remain scale-free, cross over to exponential decay, or enter a critical intermediate regime with stretched-exponential behavior. The analysis provides exact formulas for these regimes and shows that node-based deletion is less destructive to scale-free structure than ordinary random edge deletion. These results clarify how growth and degradation compete in evolving networks and identify a new route by which robust scale-free organization can be transformed into a distinct critical state.}

\section{Introduction}

 The Barabasi-Albert preferential attachment (PA) mechanism \cite{BA,AB} is probably the simplest algorithm for generating scale-free networks. Originally proposed to explain certain properties of the World Wide Web network (see also \cite{CF}), it has quickly become a foundational algorithm in the study of complex networks \cite{Newman}, with a wide range of applications. The core idea of the PA mechanism is that the newly added nodes are connected to existing ones in the network with a probability that is proportional to the degree of those existing nodes. This concept has some known precedents in the literature  \cite{Yule,Simon,Solla1,Solla2}, but the significant recent interest in this algorithm can largely be attributed to the work of Barabasi and Albert \cite{BA,AB} and their subsequent developments.
 
Many variations of the PA mechanism have been proposed and intensively investigated in the recent past. 
For instance, a growth mechanism biased by node age was considered in Ref. \cite{DM}. 
In \cite{CFV}, the authors  considered a quite general evolution
model for random networks where new nodes are added to the network  and some existing edges and nodes are deleted at each time step.
Some of the new nodes are connected to the network
according to the PA mechanism, while   others have their edges connected
uniformly to the existing nodes in the network.
  Besides this attachment  of new edges, 
   some randomly chosen, with uniform probability, nodes and edges are deleted from the network. When a node 
is deleted in this type of  evolution dynamics, all of its incident  edges are   removed.
 For further details on network evolution with node deletion, see \cite{delvert}.
Related models with   deletion of uniformly chosen edges were considered also in \cite{DeijfenLindholm,Comb},
and a case where 
  random nodes are deleted with non-uniform probability was studied in \cite{Deijfen}.
  Some mechanisms 
  with the coexistence of PA and uniform attachment rules   were investigated in \cite{Pachon}.
  For a more recent study on the network evolution with random node deletion and its relation
  with queuing theory, see \cite{Chaos}.
  In all these studies combining network growth and degradation with loss of nodes and edges, several regimes of the modified PA evolution 
 can be identified and,   typically, scale-free networks 
  with 
  power-law degree distributions with a range of exponents $\gamma>2$ can arise  asymptotically. 
These results stress, once more, the robustness and adequacy of the PA mechanism for generating and explaining scale-free networks.

In the present paper we investigate a form of edge deletion that, to our knowledge, has not been previously analyzed in this context. Specifically, we consider a growing network in which, in addition to uniform edge deletion as
discussed in \cite{Pachon}, uniformly chosen non-isolated nodes may lose one of their incident edges. This node-based deletion rule is intrinsically different from uniform edge removal.  Since a
uniformly chosen edge has a probability proportional to $k$ of terminating
in a node with degree $k$, higher degree nodes have a higher probability of losing a uniformly
removed edge. However, in many situations, the higher degree nodes are typically the most
stable ones in the network and, hence, the less likely to lose an edge. In our model, we do not have this
bias towards higher degree nodes,   all nodes
have the same probability of losing an edge. For the sake of comparison with previous work,
new nodes are added  in our model
  with the coexistence of PA and uniform attachment rules as in \cite{Pachon}, and edges
  are also uniformly deleted, as proposed originally in \cite{CFV}. 
 The resulting master equation for the stationary degree distribution becomes nonlinear through its dependence on the fraction of isolated nodes. Nevertheless, we show that the underlying nonlinear dynamics admits a complete analytical treatment and leads to three asymptotic regimes depending on the competition between attachment and deletion: scale-free (power-law), exponential, and a critical regime characterized by a
stretched exponential decay. The latter emerges from the exact balance between attachment and deletion. 
We also show that the novel node-based edge deletion mechanism
is less likely to disrupt the scale-free   regime than the usual uniform edge deletion,
further illustrating the robustness of the PA mechanism for generating  scale-free networks. Moreover, we   demonstrate that the precise balance between preferential attachment and node-based deletion can transform a scale-free network into a critical one with stretched exponential decay, a transformation impossible with the standard random edge deletion rule.  
We have performed extensive numerical simulations,  with excellent agreement with the theoretical predictions.

  We will present our model in the next Section.  Sec. III presents its analytical solutions and Sec. IV discusses the results.

\section{The model}

Our   model consists of a growing network which starts, at time step $t=0$,  from some initial random seed network with $V_0$  nodes and
$E_0$ edges.
The network then evolves 
 according to the following algorithm. At each update from $t$ to $t+1$,
\begin{enumerate}
\item with probability $p_1$, a new node is added with $m_1$ incident edges. The newly added edges are connected to the
existing  nodes at step $t$ with the usual preferential attachment rule, {\em i.e.}, the probability
of a new edge connecting to the  $i$-th node with degree $k_i$ is
\begin{equation}
\Pi_i = \frac{k_i}{2E_t},
\end{equation}
where $E_t$ stands for the total number of network edges     at time step $t$;
\item with probability $p_2 \le 1 - p_1$, a new node is added  with $m_2$ incident edges. The newly added  edges are connected to the
existing nodes at step $t$  with uniform probability;
\item with probability $p_3 \le 1 - p_1 - p_2$, $m_3$ uniformly chosen existing edges at step $t$   are deleted;
\item and, finally, with probability $p_4 = 1 - p_1 - p_2 - p_3$, $m_4$
  existing nodes with non-zero degree at step $t$ are uniformly chosen, and
one of their 
incident edges, also chosen uniformly at random, 
is deleted. 
\end{enumerate}
{ Notice that at each discrete update, exactly one of the four mutually exclusive events is selected, with $p_i\ge 0$ and 
$\sum_i p_i=1$.} The first two rules correspond  to a particular case of the Uniform-Preferential-Attachment (UPA) model  of
\cite{Pachon}. The last two rules describe a process of network degradation with the loss of edges. The fourth rule is the node-based edge-deletion mechanism introduced here.

 We focus on the network
  in the large-$t$ limit, which turns out to be independent of the details of the original seed network.  
Let $N_k(t)$ denote  the number of vertices with degree $k$ in the network  at a time step $t$. With the usual mean-field approximation (see, for instance, \cite{CFV,KR} and references therein), we obtain from the dynamics of our model  that 
\begin{eqnarray}
\label{exp}
\mathbb{E}\left[ N_k(t+1) - N_k(t)\right] &=&  p_1m_1 A_{k}(t) + p_2m_2 B_{k}(t)   +
p_3m_3 C_{k}(t) \nonumber
 \\ 
&+&    p_4m_4 D_{k}(t)  + p_1\delta^{m_1}_k + p_2\delta^{m_2}_k,
\end{eqnarray}
with $\mathbb{E}$ standing for the usual expected value of random variables and
\begin{eqnarray}
\label{A}
A_{k}(t) &=& \frac{(k-1)N_{k-1}(t)}{2E_t}   -  \frac{kN_{k}(t)}{2E_t} , \\
\label{B}
B_{k}(t) &=&  \frac{N_{k-1}(t) }{V_t}   -  \frac{N_{k}(t)}{V_t} , \\
\label{C}
C_{k}(t) &=&  \frac{(k+1)N_{k+1}(t)}{E_t}   -  \frac{kN_{k}(t)}{E_t} , \\
\label{D}
D_{k}(t) &=& \frac{  N_{k+1}(t)}{V_t^*}   -
\frac{  N_{k }(t)}{V_t^*} + \frac{(k+1)  N_{k+1}(t)}{2E_t} - \frac{k   N_{k }(t)}{2E_t} ,
\end{eqnarray}
for $k>0$
where   $V_t$ stands for 
the total     number of vertices at time step $t$, 
and $V^*_t$ is the number of 
 vertices with non-zero degree in the network  at time step $t$. For   $k=0$,
 we have
\begin{eqnarray}
\label{exp0}
\mathbb{E}\left[ N_0(t+1) - N_0(t)\right] & = & \left( \frac{2p_3m_3 + p_4m_4}{2E_t}  
+
\frac{p_4m_4}{V^*_t}
\right)N_1(t) \nonumber  \\  & -&  \frac{p_2m_2}{V_t}N_0(t). 
\end{eqnarray}
In the derivation of $D_k(t)$ above, we used that the uniform probability of choosing a
node with non-zero degree at step $t$ is $1/V^*_t$ and that the probability that
one of its incident edges terminates at a node with degree $k$ is  $k/2E_t$. The underlying construction
in this case is essentially the same one used to determine the excess degree distribution of a graph,
see \cite{Newman}, for instance. { It is important to stress that the contributions (\ref{A})-(\ref{D})  
are expected to be valid within the mean-field approximation. We will see below that we may indeed have some deviations in some particular cases.}

From the network evolution dynamics and with the usual convergence hypothesis \cite{CFV}, we have 
\begin{equation}
\label{nt}
\lim_{t\to\infty} \frac {V_t}{t} = \nu = p_1+p_2  ,
\end{equation}
and
\begin{equation}
\label{Et}
\lim_{t\to\infty} \frac {E_t}{t} = \xi = p_1m_1+p_2m_2 - p_3m_3 -p_4m_4.
\end{equation}
 Clearly,  a growing network requires $\xi > 0$. The asymptotic average degree of the network will
 be
 \begin{equation}
 \label{davg}
\left\langle k \right\rangle = \lim_{t\to\infty} \frac {2E_t}{V_t} = \frac{2\xi}{\nu}.
 \end{equation}
 The networks we are 
  concerned  here are expected to arise   asymptotically   for $t\to\infty$, when the system   attains a stationary regime. In such a regime, one can define
the probability of the network having a node with degree $k$ as
\begin{equation}
P_k = \lim_{t\to\infty} \frac{N_k(t)}{V_t},
\end{equation}
and from (\ref{nt}) we obtain 
\begin{equation}
\label{qpk}
\lim_{t\to\infty} \frac{N_k(t)}{t}  = \nu P_k
\end{equation}
and  
\begin{equation}
\lim_{t\to\infty} \frac{V_t^*}{t}  = \nu (1-P_0),
\end{equation}
since $V_t^* = V_t - N_0(t)$. Notice also that (\ref{qpk}) implies 
\begin{equation}
\lim_{t\to\infty} N_k(t+1) - N_k(t) = \nu P_k,
\end{equation}
and finally we can obtain  from   (\ref{exp})  our master equation for the network
degree distribution $P_k$  
\begin{eqnarray}
\label{Pk2}
(\alpha_2(k+2) &+& \beta_2) P_{k+2} -(\alpha_1(k+1) + \beta_1)P_{k+1} \nonumber \\
 &+& (\alpha_0 k +\beta_0) P_k = -\frac{p_1}{\nu}\delta^{m_1 }_{k+1}
 -\frac{p_2}{\nu}\delta^{m_2}_{k+1}, 
\end{eqnarray}
for $k\ge 0$, 
where 
\begin{eqnarray}
\label{coef}
\alpha_2 = \frac{  2p_3m_3 + p_4m_4}{2\xi}, \quad 
\alpha_0 = \frac{p_1m_1}{2\xi}, \quad \alpha_1 = \alpha_0+\alpha_2, \nonumber \\
\beta_2 = \frac{p_4m_4}{\nu (1-P_0)}, \quad 
\beta_0 = \frac{p_2m_2}{\nu}, \quad \beta_1 = 1+\beta_0+\beta_2.
\end{eqnarray}
Notice that   from (\ref{exp0}) we have also the following relation
between $P_0$ and $P_1$
\begin{equation}
\label{P0}
(\alpha_2 + \beta_2)P_1   - (1+\beta_0)P_0 = 0.
\end{equation}
As we will see, the uniform deletion rate being twice the node-based rate in the definition of $\alpha_2$ 
in (\ref{coef}) 
 will have significant consequences for the problem.

Before    discussing    the solutions of the master equation (\ref{Pk2}), it is worthwhile to  perform some consistency checks. 
For instance, summing both sides of    (\ref{Pk2}) and assuming that $\displaystyle \lim_{k\to\infty} kP_k = 0$,   we have
\begin{equation}
\sum_{k=0}^\infty P_k + ( \alpha_2+\beta_2)P_1 - (1+\beta_0)P_0 =1  ,
\end{equation}
which shows that, provided (\ref{P0}) holds, $P_k$ has a consistent interpretation
as a probability distribution. Moreover, multiplying both sides of (\ref{Pk2}) by $k$ and summing
again, we have
\begin{equation}
\left\langle k \right\rangle  - \frac{2\xi \left(( \alpha_2+\beta_2)P_1 - (1+\beta_0)P_0\right)}{p_1m_1 + 2p_2m_2 - p_4m_4} = \frac{2\xi}{\nu},
\end{equation}
where we have assumed that $\displaystyle\lim_{k\to\infty} k^2P_k = 0$. Again, provided (\ref{P0}) holds, we get
(\ref{davg}), confirming the consistency of our master equation (\ref{Pk2}). Let us now discuss its solutions.

\section{The solutions}
\label{sec3}

Despite its similar form to earlier recurrences, our master equation (\ref{Pk2}) is intrinsically different from  
  the class of equations  considered in \cite{CFV} or \cite{KR}, for instance. In particular, our dynamics are
non-linear, because of to the dependence of $\beta_2$ on $P_0$ and, hence, the dynamics cannot be
solved directly as an initial value problem by using the Laplace transform method of \cite{CFV}.  
Nevertheless, one can
obtain some information on the solutions of (\ref{Pk2}) for large $k$ from some simple considerations. 
Let us divide (\ref{Pk2}) by $(k+1)P_{k+1}$ and take the limit $k\to \infty$, leading to
\begin{equation}
\alpha_2 C_{k+1} -\alpha_1+ \frac{\alpha_0}{C_k} = 0,
\end{equation}
where $C_k = \frac{P_{k+1}}{P_k}$. Since $P_k$ is expected to be a probability distribution, 
  d'Alembert's ratio test implies that  
  $C_k\to c \le 1$ for large $k$, and we end up with the simple quadratic
equation
\begin{equation}
\alpha_2(c-1)\left(c - \frac{\alpha_0}{\alpha_2}\right) = 0.
\end{equation}
The $c=1$ root corresponds to power-law solutions, while  the other root $c=\frac{\alpha_0}{\alpha_2}$ will
be 
an exponentially decaying solution if $\frac{\alpha_0}{\alpha_2} < 1$. 
We can see that the parameter $\frac{\alpha_0}{\alpha_2}$  is related to the asymptotic behavior of the network degree distribution. 
Our strategy to solve the problem consists in considering $\beta_2$ a constant and search for 
  solutions of the homogeneous part of the equation (\ref{Pk2}) with the usual Laplace transform ansatz
\begin{equation}
\label{Laplace}
{P_{k+1} = \int_0^a s^k v(s) ds,}
\end{equation}
\noindent for $k\ge   m = \max\{m_1,m_2\}$,   see \cite{CFV} for further details, and then look for self-consistent solutions for the parameter $\beta_2$, which will  
eventually determine $P_0$.  
 We will  have three qualitatively different cases. We will discuss them separately below.

\subsection{The power-law case: $\alpha_0>\alpha_2$}

 {
  Inserting   (\ref{Laplace}) in 
the master equation (\ref{Pk2}) and   integrating by parts leads to
\begin{equation}
\label{LaplaceEq}
a^{k}\phi_0(a)v(a) + \int_0^a\left( s^{k-1}\phi_1(s)v(s)  - s^k\phi_0(s)v'(s)    \right)ds = 0,
\end{equation}
for $k\ge m = \max\{m_1,m_2\}$,
where
\begin{equation}
\label{phi_0}
\phi_0(s) = \alpha_2 s^2 - (\alpha_0+\alpha_2)s + \alpha_0
\end{equation}
and
\begin{equation}
\phi_1(s) = \beta_2 s^2 - (1+\beta_0+\beta_2)s + \beta_0.
\end{equation}
Notice that the roots of $\phi_0(s)$ are $s=1$ and $s=\frac{\alpha_0}{\alpha_2}$.
From (\ref{LaplaceEq}), we see that (\ref{Laplace}) will be a solution of homogeneous master equation (\ref{Pk2}) if 
\begin{eqnarray}
\label{parte}
 \phi_0(a)v(a) = 0 
\end{eqnarray}
 and
\begin{equation}
\label{eqv}
\frac{v'(s)}{v(s)} = \frac{\phi_1(s)}{s\phi_0(s)}.
\end{equation} 
Let us start with the root $a=1$ of $\phi_0(s)$.  Equation (\ref{eqv}) can be easily integrated and we will have
\begin{equation}
\label{intPK}
P_{k } = c_1 \int_0^1  s^{k+\frac{\beta_0}{\alpha_0}-1}(1-s)^{\gamma-1} \left(1-\frac{\alpha_2}{\alpha_0}s\right)^{ \lambda-1} ds,
\end{equation}
where $c_1$ is an arbitrary integration constant,
with $\gamma$ and $\lambda$ given by, respectively, 
\begin{equation}
\label{gamma}
\gamma =  1 + \frac{1}{\alpha_0 - \alpha_2} .
\end{equation}
and
\begin{equation}
\label{lambda}
\lambda = 2  -\gamma  + \frac{\beta_2}{\alpha_2} - \frac{\beta_0}{\alpha_0}.
\end{equation}
The integral (\ref{intPK}) can be expressed in terms of 
the usual integral representation of 
 hypergeometric functions, see \cite{DLMF} for instance, leading finally to 
\begin{equation}
\label{hyperPL}
P_{k } = c_1 \frac{\Gamma\left(k+\frac{\beta_0}{\alpha_0} \right)\Gamma\left(\gamma \right)}{\Gamma\left(k+\frac{\beta_0}{\alpha_0}+\gamma \right)}  {}_2F_1 \left(1-\lambda , k+\frac{\beta_0}{\alpha_0}, k+\frac{\beta_0}{\alpha_0} + \gamma   ; \frac{\alpha_2}{\alpha_0} \right) .
\end{equation} 
From the elementary asymptotic properties of the gamma function $\Gamma(z)$ and 
the asymptotic property (15.12.1) of  \cite{DLMF} for 
hypergeometric functions, it is straightforward 
to show  that  $P_k \sim k^{-\gamma}$   for large $k$.

For the 
second root $a=\frac{\alpha_0}{\alpha_2}$, we will have
\begin{equation}
\label{intQK}
Q_{k } = c_2\left(\frac{\alpha_0}{\alpha_2} \right)^{k } \int_0^1  s^{k+\frac{\beta_0}{\alpha_0}-1}(1-s)^{\lambda-1} \left(1-\frac{\alpha_0}{\alpha_2}s\right)^{ \gamma-1} ds,
\end{equation}
where $c_2$ is another integration constant. This integral can be also expressed in terms of a  hypergeometric function, leading
to 
\begin{eqnarray}
\label{hyperPLexp}
Q_{k } &=& c_2 \frac{\Gamma\left(k+\frac{\beta_0}{\alpha_0} \right)\Gamma\left(\lambda \right)}{\Gamma\left(k+\frac{\beta_0}{\alpha_0}+\lambda \right)}  \left(\frac{\alpha_0}{\alpha_2} \right)^{k } \times \\ &&   {}_2F_1 \left(1-\gamma , k+\frac{\beta_0}{\alpha_0}, k+\frac{\beta_0}{\alpha_0} + \lambda   ; \frac{\alpha_0}{\alpha_2} \right) , \nonumber 
\end{eqnarray}
Its asymptotic behavior for large $k$ can be determined analogously to the previous case, and we will have 
$Q_k \sim   k^{-\lambda}\left(\frac{\alpha_0}{\alpha_2} \right)^{k }$.  Assuming
  $\beta_2$ constant, 
the solutions $P_k$ and $Q_k$ are the two linearly independent solutions of (\ref{Pk2}),
for $k\ge m = \max\{m_1,m_2\}$, as one can see from the Casoratian determinant \cite{book}
\begin{equation}
\label{casorat}
\mathcal{C}_k =
\left| 
\begin{array}{cc}
P_k & Q_k \\
P_{k+1} & Q_{k+1}
\end{array}
\right|
 ,
\end{equation}
which   for large $k$ is such that 
\begin{equation}
\label{casot-asympt}
\mathcal{C}_k \sim k^{-\lambda - \gamma } \left(\frac{\alpha_0}{\alpha_2} \right)^{k } > 0,
\end{equation} 
 For the present case with $\frac{\alpha_0}{\alpha_2} > 1$, the solution $Q_k$ diverges for $k\to\infty$ and, hence, it cannot describe a valid probability distribution, leading to $c_2=0$ and, consequently, 
 the degree distribution $P_k$ will be given by  (\ref{hyperPL}). 
 }

  We 
have 
 two constants to be determined in order to complete the solution of our problem, namely the integration constant $c_1$   and the value of $P_0$. 
 They obey a non-linear relation due to the  dependence of $\beta_2$, and consequently of $\lambda$, on $P_0$, in 
(\ref{coef}) and 
 (\ref{hyperPL}).
In order to determine $P_0$, 
let us inspect the quantity $C_k = \frac{P_{k+1}}{P_k}$, which, by construction, does not depend on $c_1$
for $k\ge m$. We have
\begin{eqnarray}
\label{plC}
C_k &=& h_k(P_0) \\ 
&=& \frac{k+\frac{\beta_0}{\alpha_0}}{k+\frac{\beta_0}{\alpha_0}+ \gamma}
\frac{{}_2F_1 \left(1-\lambda , k+\frac{\beta_0}{\alpha_0}+1, k+\frac{\beta_0}{\alpha_0} + \gamma+1   ; \frac{\alpha_2}{\alpha_0} \right)}{{}_2F_1 \left(1-\lambda , k+\frac{\beta_0}{\alpha_0}, k+\frac{\beta_0}{\alpha_0} + \gamma   ; \frac{\alpha_2}{\alpha_0} \right)}. \nonumber
\end{eqnarray}
 On the other hand, for moderate values of $k$, $C_k$ can be easily evaluated 
by simply   iterating (\ref{Pk2}) and (\ref{P0}). Let $C_k = g_k(P_0)$ be the rational function obtained in this way.  
The value of
$P_0$ can be (numerically) determined from the non-linear continuity condition $g_m(P_0)=h_m(P_0)$. Once $P_0$ is determined,
the integration constant $c_1$ can be set from the value of $P_{m}$ evaluated iteratively from (\ref{Pk2}), and the problem will be completely solved. For practical purposes, we expect small values for $P_0$, so we can consider as a first approximation
the relation $g_m(P_0)=h_m(0)$, which yields  a polynomial equation for $P_0$, and then refine the solution by exploring the
nonlinear function $h_m(P_0)$.

{
It is also instructive to inspect the corresponding survival (complementary cumulative) degree distributions 
\begin{equation}
\label{survival}
S_k = \sum_{j=k}^\infty P_j
\end{equation}
for our evolving networks. From (\ref{intPK}) we have
\begin{equation}
S_k  =  c_1 \int_0^1  s^{k+\frac{\beta_0}{\alpha_0}-1}(1-s)^{\gamma-2} \left(1-\frac{\alpha_2}{\alpha_0}s\right)^{ \lambda-1} ds,
\end{equation}
which can be also expressed in terms of a hypergeometric function. More specifically, it corresponds to the same expression in equation (\ref{hyperPL}), but with the substitution $\gamma \to \gamma - 1$. Notably, its asymptotic behavior for large $k$ will be $S_k \sim k^{1-\gamma}$.
}

\begin{figure*}[ht]
\includegraphics[width=0.49\linewidth]{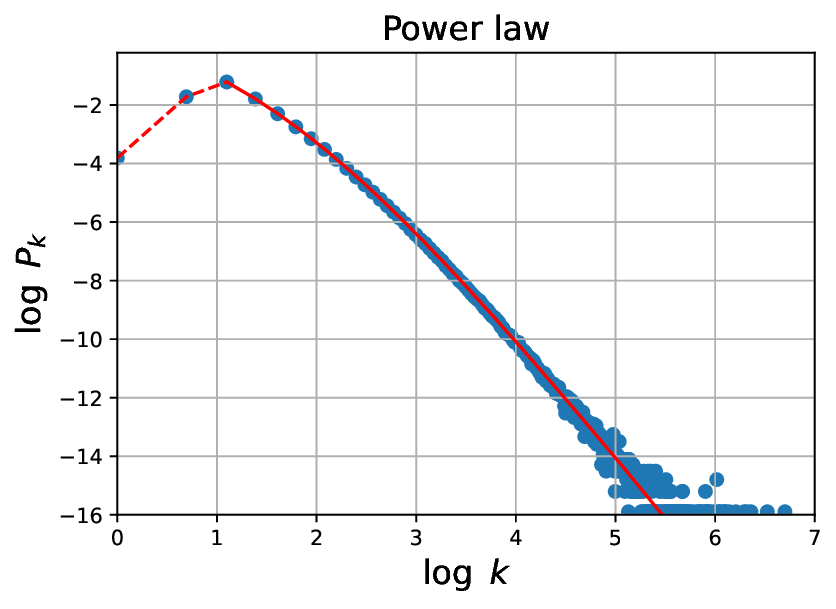}
\includegraphics[width=0.49\linewidth]{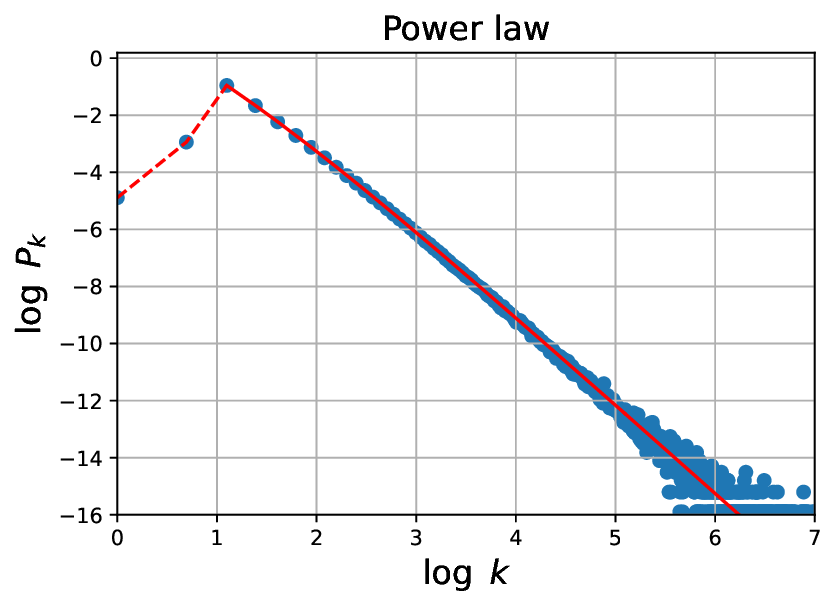} 
\includegraphics[width=0.49\linewidth]{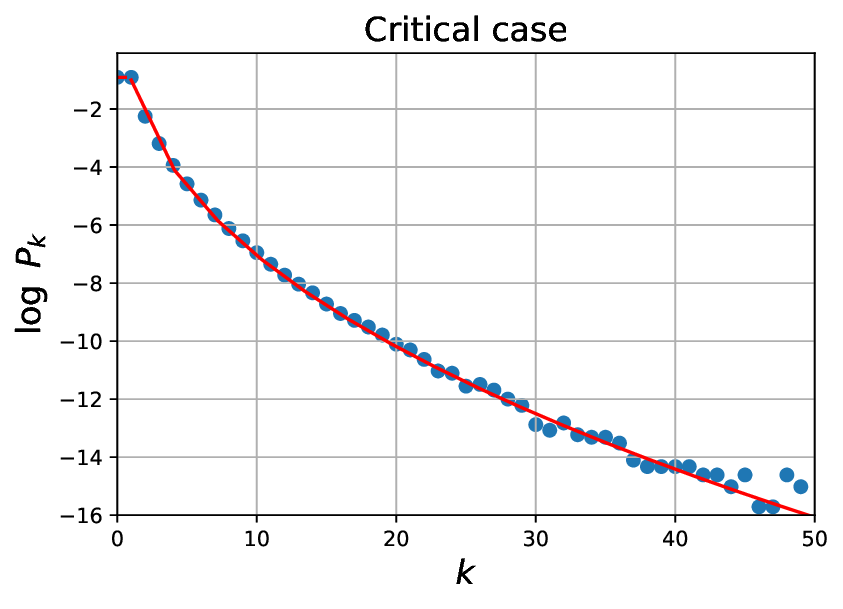} 
\includegraphics[width=0.49\linewidth]{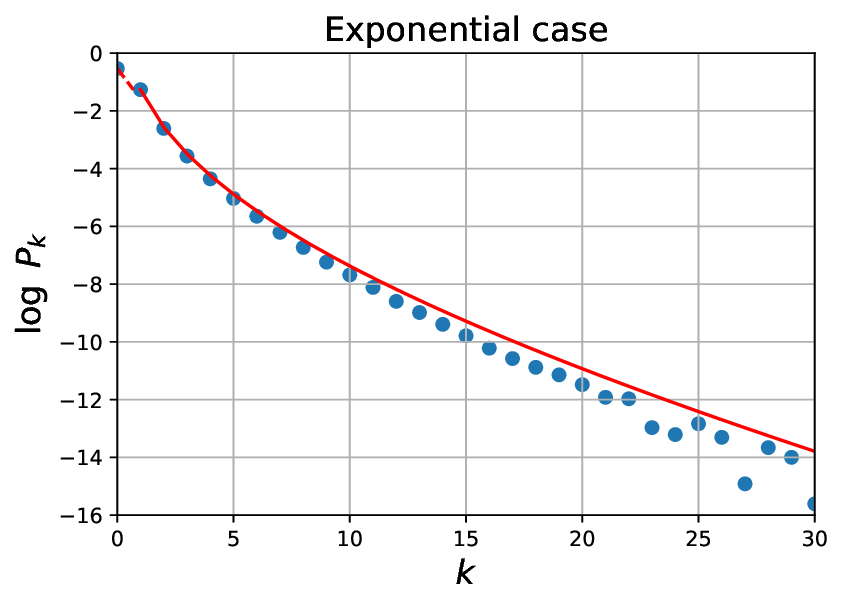}  
\caption{\label{fig1}  Degree distributions for random networks obtained by running $10^7$
steps of our algorithm (blue dots) and the theoretical predictions in terms of hypergeometric functions (solid red line),
{ which were obtained from (\ref{Laplace}) for $k\ge m =\max\{m_1,m_2\}$. 
The theoretical prediction for $0\le k < m$ (traced red line) are obtained from the iteration of (\ref{Pk2}) and
(\ref{P0}), starting from the theoretical value for $P_0$.}
Top left: Power-law case
with all rules of our algorithm, 
 corresponding to $p_1=0.5$, $p_2=0.3$, $p_3 = p_4 = 0.1$, $m_1=3$,  
$m_2=2$, and $m_3=m_4=1$. 
From the theoretical predictions, we have $P_0\approx 2.63\times 10^{-3} $, while from the simulation
we obtained $P_0\approx 2.51\times 10^{-3} $. Top right: Another power-law case 
with PA and the two deletion rules, 
corresponding to $p_1=0.8$, $p_2=0$, $p_3 = p_4 = 0.1$, and $m_1 =3$ and $m_3=m_4=1$. 
The theoretically predicted and the observed values for $P_0$ in this case are
$1.40\times 10^{-3} $ and $1.38\times 10^{-3} $, respectively.
Bottom left: Critical case with PA and uniform edge deletion, corresponding to $p_1=2/3$, $p_3=1/3$, $p_2 = p_4 = 0$, and $m_1=m_3=1$.  
Since we do not have the node-based edge deletion in this case ($\beta_2=0$), we do not need to solve the nonlinear equation
to determine $P_0$.
Bottom right: Exponential case with PA and the two deletion rules, corresponding to $p_1=0.6$, $p_2=0$, $p_3 = 0.3$, $ p_4 = 0.1$, and $m_1=m_3=m_4=1$.  The theoretically predicted and observed values for $P_0$  given by, respectively, $5.80\times 10^{-1} $ and $5.89\times 10^{-1} $.
In all panels, the red lines correspond  to the respective theoretical  prediction (\ref{hyperPL}), (\ref{hyperPLexp}), or (\ref{kummer}) for $k\ge m = \max\{m_1,m_2\}$. { It is important to stress that our model is such that, at each discrete update, exactly one of the four mutually exclusive events is selected, with $p_i\ge 0$ and 
$\sum_i p_i=1$. The simulations and the master equation (\ref{Pk2}) use this same convention.}
 We obtained very good overall agreement for all considered cases. It is worth   noticing that for the last two cases, namely the sub-exponential and the exponential decaying cases, it is intrinsically more expensive computationally to achieve good statistical convergence. 
{ We have identified a systematic bias for the exponential case, namely that 
the simulated tails falls below the theoretical mean-field prediction. We will return to this issue in the Discussion section.
} 
 For all cases we have considered, the nonlinear equations for $P_0$ based on (\ref{plC}), (\ref{expC}), and (\ref{ccC}) had a unique solution. 
}
\end{figure*}

\begin{figure*}[ht]
\includegraphics[width=0.49\linewidth]{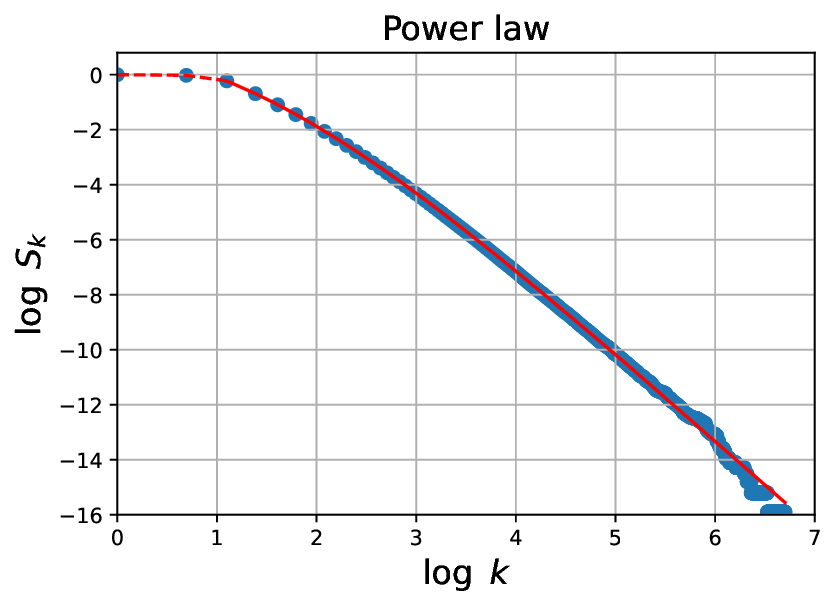}
\includegraphics[width=0.49\linewidth]{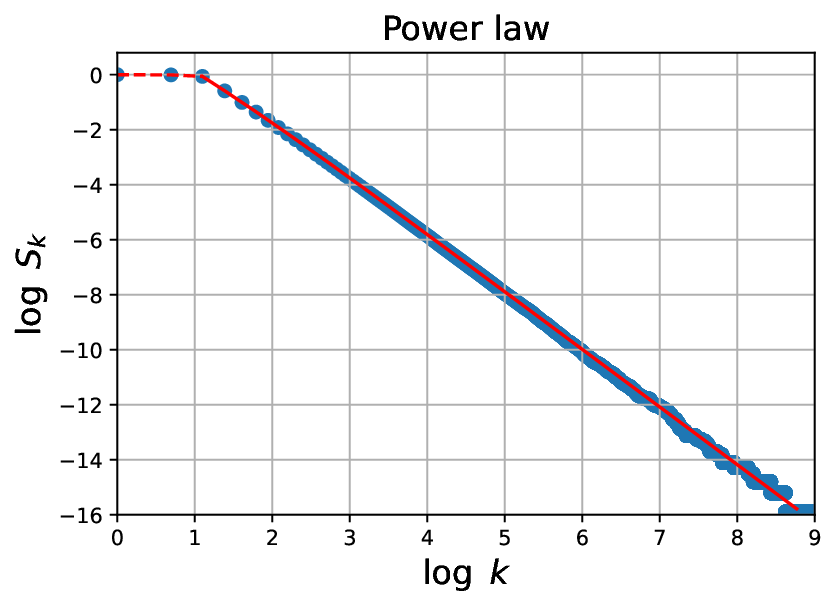} 
\includegraphics[width=0.49\linewidth]{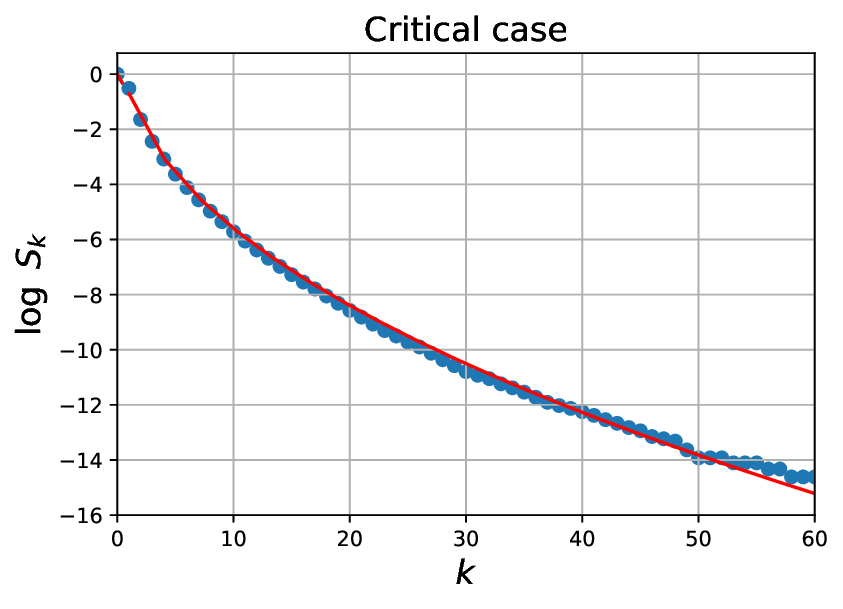} 
\includegraphics[width=0.49\linewidth]{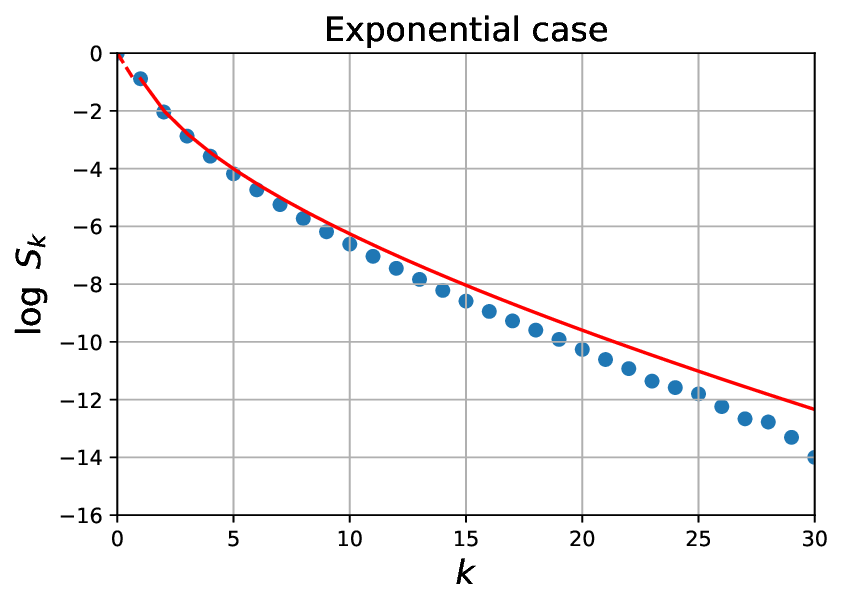}  
\caption{\label{fig2}  { Survival
(complementary cumulative) degree distributions for the cases considered in Fig. \ref{fig1}, with the same parameters
and conventions.  
It is clear that the simulated tails falls below the theoretical mean-field prediction  for the the exponential case. 
 We will return to this issue in the last section.
}
}
\end{figure*}

We   have checked our solutions against comprehensive numerical simulations. 
Figs. \ref{fig1} and \ref{fig2} summarize   some typical  cases obtained by running $10^7$
steps of our algorithm. We obtained very good overall agreement for all
considered cases.
 
\subsection{The exponential case: $\alpha_0<\alpha_2$}

{ The solutions of the previous case are formally valid also for the exponential case $\frac{\alpha_0}{\alpha_2} < 1$. However, the asymptotic behavior for the solutions (\ref{hyperPL}) and (\ref{hyperPLexp}) are reversed, and
we need to impose $c_1=0$ in order to get valid
probability distribution solutions, leading   to a degree distribution $P_k$ for the exponential case given by
the expression (\ref{hyperPLexp}),} which decays as $P_k \sim   k^{-\lambda}\left(\frac{\alpha_0}{\alpha_2} \right)^{k }$ for large $k$, with $\lambda$ given by (\ref{lambda}).
 The values of
$P_0$ and $c_2$ can be determined analogously to those in the previous power-law case.
The function $h_m(P_0)$ in this case reads
\begin{equation}
\label{expC}
 h_m(P_0) = \frac{m'}{m'+ \lambda} \frac{\alpha_0}{\alpha_2}
\frac{{}_2F_1 \left(1-\gamma , m'+1, m' + \lambda+1   ; \frac{\alpha_2}{\alpha_0} \right)}{{}_2F_1 \left(1-\gamma , m', m' + \lambda   ; \frac{\alpha_2}{\alpha_0} \right)},
\end{equation}
where $m'= m+ \frac{\beta_0}{\alpha_0}$.
{ The corresponding survival  degree distributions $S_k$ in this case can
be determined analytically in the same way of the previous power-law case, and the result will also be the expression
(\ref{hyperPLexp}) with the substitution $\gamma \to \gamma -1$. In this case, the asymptotic behavior is  unaltered,
$S_k \to k^{-\lambda}\left(\frac{\alpha_0}{\alpha_2} \right)^{k } $ for large $k$.}
 A typical situation for the exponential case is
  depicted in Figs. \ref{fig1} and \ref{fig2},  also for a simulation with $10^7$
steps of our algorithm.

\subsection{The critical case}
{

Finally, we are left with the critical case, which is
rather more intricate and has not been explored in depth in the literature yet.  The first observations is that, in this case, the polynomial  (\ref{phi_0}) is $\phi_0(s)=\alpha_0(s-1)^2$, and the integration
of (\ref{eqv}) for the double root $a=1$ leads to the solution 
\begin{equation}
\label{criticaleq0}
P_k = c_1 \int_0^1 s^{k+\frac{\beta_0}{\alpha_0}-1}(1-s)^\frac{\beta_2-\beta_0}{\alpha_0}\exp\left({-\frac{ 1}{\alpha_0 (1-s)}}\right)ds.
\end{equation}
In terms of the new integration variable $r = \frac{s}{1-s}$, we have
\begin{equation}
\label{criticaleq}
P_k = c_1   \int_0^\infty r^{k+\frac{\beta_0}{\alpha_0}-1}(r+1)^{k+\frac{\beta_2}{\alpha_0}+1}
e^{-\frac{r}{\alpha_0}}
dr,
\end{equation}
where the constant $e^{-\frac{1}{\alpha_0}}$ was absorbed in $c_1$. The expression (\ref{criticaleq}) 
is an integral representation  for the Kummer's confluent function of the second kind (Tricomi function), see 
 (13.4.4)
of \cite{DLMF}, 
  leading
 to 
 \begin{equation}
\label{kummer}
P_k = c_1 \Gamma\left(k+\frac{\beta_0}{\alpha_0}\right)U\left(k+\frac{\beta_0}{\alpha_0},\frac{\beta_0-\beta_2}{\alpha_0},\frac{1}{\alpha_0}\right).
\end{equation} 
 From the asymptotic behaviors 
 (13.8.11) and (10.40.2)  of \cite{DLMF}, corresponding, respectively, to the
 of the Tricomi and modified Bessel functions,   
  one can show that 
   $P_k\sim  k^{-\delta}e^{-2\sqrt{\frac{k}{\alpha_0}}}$ for large $k$, where $\delta$ is given by 
\begin{equation}
\label{delta}
\delta = \frac{3}{4} + \frac{\beta_2-\beta_0}{2\alpha_0}.
\end{equation}   
   We need now to show
   that the second solution $Q_k$ in the critical case will not correspond to a probability distribution. This can be done
   by invoking 
 Heymann's theorem \cite{book} for the Casoratian determinant (\ref{casorat}),
which implies in the present case  the following first order difference equation  
\begin{equation}
\left(\alpha_2 (k+2) + \beta_2\right)\mathcal{C}_{k+1} = \left(\alpha_0k + \beta_0\right)\mathcal{C}_{k},
\end{equation}
whose solution is
\begin{equation}
\mathcal{C}_k= c_3 
\frac{\Gamma\!\left(k+\frac{\beta_0}{\alpha_0}\right)}{\Gamma\!\left(k+2+\frac{\beta_2}{\alpha_2}\right)} ,
\end{equation}
where $c_3$ is a non-vanishing integration constant, from which  we recover (\ref{casot-asympt}) for $\alpha_0=\alpha_2$, implying that
the second solution $Q_k$ for the critical case grows sub-exponentially for large  $k$ and, hence, does not admit  an interpretation as a probability distribution.}

Thus, in the critical case, the degree distribution exhibits am asymptotic stretched-exponential decay, which is slower than exponential but faster than any power law. This regime arises from the exact compensation between attachment and deletion processes. 
The values of
$P_0$ and $c_1$ can be determined analogously as to the previous case, with 
the function $h_m(P_0)$ in this case given by
\begin{equation}
\label{ccC}
h_m(P_0) = \left( m+ \frac{\beta_0}{\alpha_0} \right) \frac{U\left(m+\frac{\beta_0}{\alpha_0}+1,\frac{\beta_0-\beta_2}{\alpha_0},\frac{1}{\alpha_0}\right)}{U\left(m+\frac{\beta_0}{\alpha_0},\frac{\beta_0-\beta_2}{\alpha_0},\frac{1}{\alpha_0}\right)}.
\end{equation}
{ As to the survival distribution, we have from (\ref{criticaleq0}) 
\begin{equation}
S_k = c_1 \int_0^1 s^{k+\frac{\beta_0}{\alpha_0}-1}(1-s)^{\frac{\beta_2-\beta_0}{\alpha_0}-1}\exp\left({-\frac{ 1}{\alpha_0 (1-s)}}\right)ds,
\end{equation}
which can be expressed in terms of the same Kummer function (\ref{kummer} with the substitution $\frac{\beta_2}{\alpha_0}\to \frac{\beta_2}{\alpha_0}-1 $. The corresponding asymptotic behavior will be $S_k\sim  k^{\frac{1}{2}-\delta}e^{-2\sqrt{\frac{k}{\alpha_0}}}$ for large $k$, with $\delta$ given by (\ref{delta}).}
A
 typical situation for the critical case with  $10^7$
steps of our algorithm is depicted in Figs. \ref{fig1} and \ref{fig2}.

\section{Discussion}

We have  introduced and solved a generalized evolving network model combining mixed attachment mechanisms with two distinct forms of edge deletion, particularly including a node-based deletion process that has not been previously considered. The resulting dynamics are intrinsically nonlinear, yet admit  a complete  classification of degree distributions. Our main result can be summarized
in the following three distinct asymptotic   classes:
\begin{equation}
\label{solut}
P_k \sim \left\{ \displaystyle 
\begin{array}{ll}
k^{-\gamma}, & \quad \alpha_0 > \alpha_2, \\ 
 \displaystyle  k^{-\delta}e^{-2\sqrt{\frac{k}{\alpha_0}}}, & \quad \alpha_0 = \alpha_2 ,\\ 
k^{-\lambda}\left(\frac{\alpha_0}{\alpha_2} \right)^k, & \quad \alpha_0 <  \alpha_2 ,
\end{array}
\right.
\end{equation}
with the exponents $\gamma$, $\delta$, and $\lambda$   given, respectively, by
(\ref{gamma}), (\ref{delta}), and (\ref{lambda}). { In terms of the 
survival (complementary cumulative) degree distributions $S_k$, the classes are
\begin{equation}
\label{solut1}
S_k \sim \left\{ \displaystyle 
\begin{array}{ll}
k^{1-\gamma}, & \quad \alpha_0 > \alpha_2, \\ 
 \displaystyle  k^{\frac{1}{2}-\delta }e^{-2\sqrt{\frac{k}{\alpha_0}}}, & \quad \alpha_0 = \alpha_2 ,\\ 
k^{-\lambda}\left(\frac{\alpha_0}{\alpha_2} \right)^k, & \quad \alpha_0 <  \alpha_2 .
\end{array}
\right.
\end{equation}}
According to the definitions (\ref{coef}), the parameter $\alpha_0$
 quantifies the strength of preferential attachment due to the addition of new nodes, whereas 
 $\alpha_2$ 
 measures the effective rate of edge loss induced by both uniform edge deletion and node-based edge removal.
Interestingly, from the definition of $\alpha_2$, one can see   that uniform edge deletion is twice as likely to disrupt the scale-free (power-law) regime compared to node-based edge deletion. 
We have   checked the solutions (\ref{solut}) and (\ref{solut1}) against comprehensive numerical simulations. 
Figs. \ref{fig1} and \ref{fig2}  depict some degree and survival distributions    for networks obtained by running $10^7$
steps of our algorithm.  We obtained very good overall agreement for all
considered cases. { However, we still have the issue of the simulated tails falling below the theoretical mean-field prediction for the exponential case. We conducted simulations using different total numbers of steps, and we found that the bias towards lower degrees persists for this set of parameters related to the exponential case, indicating that poor statistics and finite-size effects are unlikely to be the main reason for the observed systematic mismatch. The probable cause is a failure of the mean-field approximation for this particular parameter set. Since the mismatch is stronger when our node-based edge deletion process is present, this is likely related to the probability $D_k(t)$, whose mean-field approximation is given by (\ref{D}). In the derivation of such a probability in the mean-field approximation, we used that the uniform probability of choosing a
node with non-zero degree at step $t$ is $1/V^*_t$ and that the probability that one of its incident edges terminates at a node with degree $k$ would be $k/2E_t$. However, we should  not expected that the latter be a good approximation for networks with lower average degrees. For instance, for the exponential case depicted in Figs. \ref{fig1} and  \ref{fig2}, the average degree, given by (\ref{davg}), is $\langle k\rangle = \frac{2}{3}$. In such networks with $\langle k\rangle < 1$, we do not expect the probability that an edge terminates at a node with degree $k$ to be $k/2E_t$, simply because nodes with degree $k>1$ are extremely rare for these networks and typically connected only to $k=1$ nodes (leaves).
 In these situations, to which our exponential case belongs, our mean-field analysis tends to overestimate the contribution of higher-degree nodes, and the net result will be a theoretical mean-field prediction above the simulation tails, exactly as observed in Figs. \ref{fig1} and \ref{fig2}. This is certainly not a definitive answer to this question,  which lies  beyond the scope of the present work. This is still under investigation. }  

The asymptotic classes (\ref{solut}) and  (\ref{solut1}) 
    provide a complete physical interpretation of the growth dynamics of the 
    evolving network. Its  asymptotic topology  is governed by the competition between the network growth and degradation
    by losing edges.
  When preferential attachment dominates ($\alpha_0 > \alpha_2$), the network retains a scale-free structure with a tunable power-law exponent. { Taking into consideration (\ref{coef}), the power-law exponent (\ref{gamma}) can be written
  as
\begin{equation}
\label{gamma1}
\gamma = 3 + 2\frac{p_2m_2+p_3m_3}{p_1m_1 - 2p_3m_3 - p_4m_4} \ge 3.
\end{equation}  
Thus, our model produces power-law networks with exponent $3\le \gamma < \infty$, approaching infinity at the critical boundary $\alpha_0 = \alpha_2$. Our model is not able to access the range $2< \gamma < 3$. In the pure preferential-attachment plus node-based-deletion case, $p_2=p_3=0$, one obtains $\gamma=3$ throughout the entire growing power-law regime.} On the other hand, when the network degradation dominates, 
high-degree vertices are increasingly suppressed and the distribution becomes exponentially decaying. These two regimes are consistent with earlier studies of growing networks with edge or node deletion,
  see for instance \cite{CFV},  although the quantitative exponents differ due to the node-based deletion mechanism introduced here.
  
However,  the most notable outcome here is the critical regime $\alpha_0=\alpha_2$. In this balanced case, the degree distribution exhibits a stretched exponential decay, decaying more slowly than exponential but faster than any power law. One can see that this regime  
 emerges from the balance between global growth and decay rates, without additional fine-tuning,  suggesting that it may be robust in realistic systems where expansion and degradation naturally counterbalance. 
Stretched-exponential and other sub-exponential
 degree distributions have been repeatedly observed in empirical studies of diffusion processes, including information spreading, cultural adoption, and language dynamics. In particular, recent large-scale analyses of online word diffusion have shown that usage frequencies often grow sub-exponentially over time, with some growth curves well described by  stretched-exponential forms, see
 for instance
  \cite{str1,str2,Watanabe}. Although the objects of interest in these studies are different,
  the underlying mechanisms share a common feature, namely a balance between network growth and degradation.
 This suggests that the stretched-exponential degree distributions arising in the critical case might be particularly relevant for modeling real-world networks that sustain long-term growth with random loss of edges.
 
 { In summary, the three stationary mean-field regimes can also be expressed
directly in terms of the original model parameters, as depicted in
Table~\ref{tab:phase-regimes}.
\begin{table}[t]
\caption{Stationary mean-field regimes of the model. All cases assume
the growing-network condition
$p_{1}m_{1}+p_{2}m_{2}>p_{3}m_{3}+p_{4}m_{4}$.}
\label{tab:phase-regimes}
\centering
\begin{tabular}{lc}
\hline\hline
Regime & Condition \\[2pt]
\hline
Power law
&
$p_{1}m_{1}>2p_{3}m_{3}+p_{4}m_{4}$
\\[2pt]
Critical (stretched exponential)
&
$p_{1}m_{1}=2p_{3}m_{3}+p_{4}m_{4}$
\\[2pt]
Exponential
&
$p_{1}m_{1}<2p_{3}m_{3}+p_{4}m_{4}$
\\[2pt]
\hline\hline
\end{tabular}
\end{table}
The table shows that the uniform-attachment rate $p_{2}m_{2}$ does
not enter the phase boundary directly, although it contributes to
$\xi$ and therefore affects the exponents, the normalization, and the
fraction of isolated vertices.  }

 Finally, it is worth noting that in situations where only PA and the node-based edge deletion are present, the condition of a growing network requires $p_1m_1 > p_4m_4$, and, consequently, all asymptotic networks arising from our algorithm will be of the scale-free type, with $\alpha_0>\alpha_2$. However, if we start with a large scale-free network and run our algorithm with $p_2 = p_4= 0$ and $p_1m_1 = 2p_3m_3$, after a large number of iterations, we will ultimately arrive at a stretched exponential decaying network,  which would not be achievable if we had started with PA and the node-based   edge deletion. This degradation mechanism for scale-free networks, as well as some possible applications, is now under investigation.

\section*{Acknowledgment}
This work was conducted in partial fulfillment of the requirements for the Master's degree of E.R.C. at the University of Campinas.
A.S. is partially       supported by CNPq (Brazil) grant 306785/2022-6 
and wishes to thank Profs. Vitor Cardoso and Jose S. Lemos for the warm hospitality
at the Center for Astrophysics and Gravitation of the
University of Lisbon, where part of this work was done. { Finally, we thank the anonymous referees for raising interesting points that have significantly improved our paper.  }


\begin{thebibliography}{99}

\bibitem{BA} A.L. Barabasi  and R. Albert, Science
{\bf 286}, 509 (1999).

\bibitem{AB} R. Albert  and A.L. Barabasi,  Rev.
Mod. Phys. {\bf 74}, 47  (2002).

\bibitem{CF}C. Cooper  and  A. Frieze, Random Struct. Alg. {\bf 22},  311 (2003).

\bibitem{Newman} M. Newman, {\em Networks: an Introduction}, Oxford University Press (2010).




\bibitem{Yule} G.U. Yule, Philos. Trans. R. Soc. B {\bf 213}, 21 (1925).

\bibitem{Simon} H. A. Simon,  Biometrika {\bf 42}, 425 (1955).

\bibitem{Solla1} D.J. de Solla Price,  Science {\bf 149}, 510 (1965).

\bibitem{Solla2} D.J. de Solla Price,  J. Amer. Soc. Inform. Sci. {\bf 27}, 292 (1976).

\bibitem{DM} S. N. Dorogovtsev and J. F. F. Mendes,
Phys. Rev. E {\bf 62}, 1842 (2000) [arXiv:cond-mat/0001419]

\bibitem{CFV} C. Cooper, A. Frieze, and J. Vera, Internet Mathematics {\bf 1},
463 (2004).

\bibitem{delvert} C. Moore, G. Ghoshal, and M. E. J. Newman, Phys. Rev. E {\bf 74}, 036121 (2006). [arXiv:cond-mat/0604069]

\bibitem{DeijfenLindholm} M. Deijfen and M. Lindholm, Physica A 3{\bf 88}, 4297  (2009). [arXiv:1509.07032]


\bibitem{Comb} T. DuBois, S. Eubank,  and A. Srinivasan, Electron. J. Comb. {\bf 19},
P51 (2012).

\bibitem{Deijfen} M. Deijfen, J.  Appl. Prob. {\bf 47}, 1150  (2010). [arXiv:1509.07033]



\bibitem{Pachon}A. Pachon, L. Sacerdote, and S. Yang, Physica D {\bf  371}, 1 (2018).
[arXiv:1704.08597]



\bibitem{Chaos} M. Feng, L. Deng, and K. Kurths, Chaos {\bf 28}, 083118 (2018).


\bibitem{KR} P. L. Krapivsky and S. Redner,
Phys. Rev. E {\bf 63}, 066123 (2001) [arXiv:cond-mat/0011094].


\bibitem{DLMF} 
F. W. J. Olver, A. B. Olde Daalhuis, D. W. Lozier, B. I. Schneider, R. F. Boisvert, C. W. Clark, B. R. Miller, B. V. Saunders, H. S. Cohl, and M. A. McClain, (Eds.), NIST Digital Library of Mathematical Functions. \href{https://dlmf.nist.gov/}{\tt https://dlmf.nist.gov/}. 


\bibitem{book}  L.M. Milne-Thomson, {\em  Calculus of Finite Differences}, (Reprinted) AMS Chelsea Publishing, Providence RI
(1980).

\bibitem{str1} { A.D. Broido and A. Clauset, Nature Comm. {\bf 10}, 1017 (2019). [arXiv:1801.03400]}

\bibitem{str2} { G. Ghoshal, L. Chi, and  A.L. Barabasi, Scientific Reports {\bf 3},  2920 (2013).}

 
 \bibitem{Watanabe} H. Watanabe, Phys. Rev. E {\bf 114}, 014304 (2026). [arXiv:2511.04106]  






 






\end{thebibliography}
 \end{document}